\documentclass[conference,letterpaper]{IEEEtran}

\usepackage{cite}
\usepackage{amsmath,amssymb,amsfonts}
\usepackage[ruled]{algorithm2e}         
\usepackage{graphicx}
\usepackage{textcomp}
\usepackage{xcolor}
\usepackage{upgreek}
\usepackage{siunitx}
\usepackage[caption=false,font=footnotesize]{subfig}
\usepackage{booktabs}
\usepackage{tabularx}
\usepackage{tikz}
\usetikzlibrary{matrix,fit}
\pgfkeys{tikz/mymatrix/.style={matrix of math nodes,inner sep=0pt,row sep=0em,column sep=0em,nodes={inner sep=6pt}}}

\def\BibTeX{{\rm B\kern-.05em{\sc i\kern-.025em b}\kern-.08em
    T\kern-.1667em\lower.7ex\hbox{E}\kern-.125emX}}

\usepackage[hidelinks]{hyperref}

\begin{document}

\title{Forecasting Source Stability in Scientific Experiments using Temporal Learning Models: A Case Study from Tritium Monitoring\\

}


\author{
\IEEEauthorblockN{
Nicholas Tan Jerome\IEEEauthorrefmark{1},
Nadia Aouadi\IEEEauthorrefmark{2},
Christoph Köhler\IEEEauthorrefmark{3}\IEEEauthorrefmark{4},\\
Suren Chilingaryan\IEEEauthorrefmark{1},
Andreas Kopmann\IEEEauthorrefmark{1}
}
\IEEEauthorblockA{\IEEEauthorrefmark{1}Institute for Data Processing and Electronics, Karlsruhe Institute of Technology, Germany\\
\{nicholas.tanjerome, suren.chilingaryan, andreas.kopmann\}@kit.edu}
\IEEEauthorblockA{\IEEEauthorrefmark{2}Schmalkalden University of Applied Sciences, Schmalkalden, Germany\\
nadiaaouadi.contact@gmail.com}
\IEEEauthorblockA{\IEEEauthorrefmark{3}Technical University Munich, Garching, Germany\\
\IEEEauthorrefmark{4}Max-Planck-Institut für Kernphysik, Heidelberg, Germany\\
christoph.koehler@tum.de}
\IEEEauthorblockA{\textit{Author note:} Work by Nadia Aouadi was performed during an internship at IPE, KIT.}

}

\maketitle

\begingroup
\renewcommand\thefootnote{}\footnote{
This is the author's accepted manuscript. The final published version is available at: https://doi.org/10.1109/ICDMW69685.2025.00038. © 2025 IEEE.
}
\addtocounter{footnote}{-1}
\endgroup

\begin{abstract}
The Karlsruhe Tritium Neutrino Experiment (KATRIN) aims to measure the absolute neutrino mass with unprecedented sensitivity, requiring precise monitoring of the windowless gaseous tritium source, where tritium beta decay occurs. To track variations of the source activity, beta-induced X-ray spectroscopy provides real-time diagnostics. However, traditional drift detection methods struggle with the infrequent and transient nature of instability events in gaseous tritium. This study bridges the gap between state-of-the-art time-series forecasting models and real-world experimental applications by leveraging deep learning to predict the time to stability after instabilities. Unlike standard benchmarking approaches that emphasize algorithmic performance on fixed datasets, we apply forecasting models—including LSTM, N-BEATS, TFT, NHITS, DLinear, NLinear, TSMixer, and Chronos-LLM—to complex, large-scale experimental data. Our findings highlight two challenges: learning from sparse instability events and forecasting long time horizons (i.e., predicting hundreds of future points), both of which are ongoing challenges in time-series forecasting and remain active areas of research. This prediction task has direct experimental value by enabling better scheduling and maintenance planning. A reliable forecast of stability time allows for more efficient measurement and task management during stabilization periods. Through model selection, we identified N-BEATS as the top performer, excelling in accuracy and repeatability, demonstrating that deep learning can optimize large-scale physics experiments.
\end{abstract}


\section{Introduction}

Neutrinos are fundamental particles whose exact mass remains unknown. Pinning down their mass could unlock some of the universe’s deepest mysteries~\cite{nature2022}. The Karlsruhe Tritium Neutrino (KATRIN) experiment aims to measure the absolute neutrino mass scale with unprecedented sensitivity. Measurements began in 2019 and will continue until the end of 2025~\cite{nature2019,parno2022}. A key component of the 70-meter-long experiment is the Windowless Gaseous Tritium Source (WGTS), a stable and an ultra-luminous $\upbeta$-decay source that delivers up to 
$10^{11}$ decays per second by circulating high-purity molecular tritium. The WGTS consists of a \SI{10}{\meter}-long, \SI{90}{\milli\meter}-diameter tube, where tritium gas is continuously injected at the center and diffuses toward both ends before being recaptured and pumped away. Maintaining a steady gas flow and stable tritium concentration requires dedicated loop and buffer systems for supply and purification. 

However, short-term instabilities can arise from temperature fluctuations in the WGTS, which affect conductance and disrupt the gas flow. Additionally, operation of the tritium source with nominal gas density is periodically interrupted for maintenance and calibration procedures that require an empty source. Variations in the circulation loop restart arise from different procedures, system states, and operator actions. If the source operation is interrupted only briefly, the gas injection system is closed while tritium remains in the loop, and restarting simply involves reopening the gas injection system. When restarting from a fully evacuated state, the loop is first prefilled to a level determined by the operator, and opening the gas injection system may cause gradual or abrupt refills. To reduce initial overshoot, operators may adjust the gas flow settings. In some cases, restarting too quickly can trigger an interlock safety control mechanism, leaving the system in an unusual state. This, the stabilization process depends on operational history and operator decisions. After refilling, it takes up to several hours for the WGTS to return to a stable equilibrium~\cite{aker2021design}. Ideally, neutrino mass measurements resume as soon as stability is restored.

\begin{figure*}[t]
  \centering
      \includegraphics[width=0.95\textwidth]{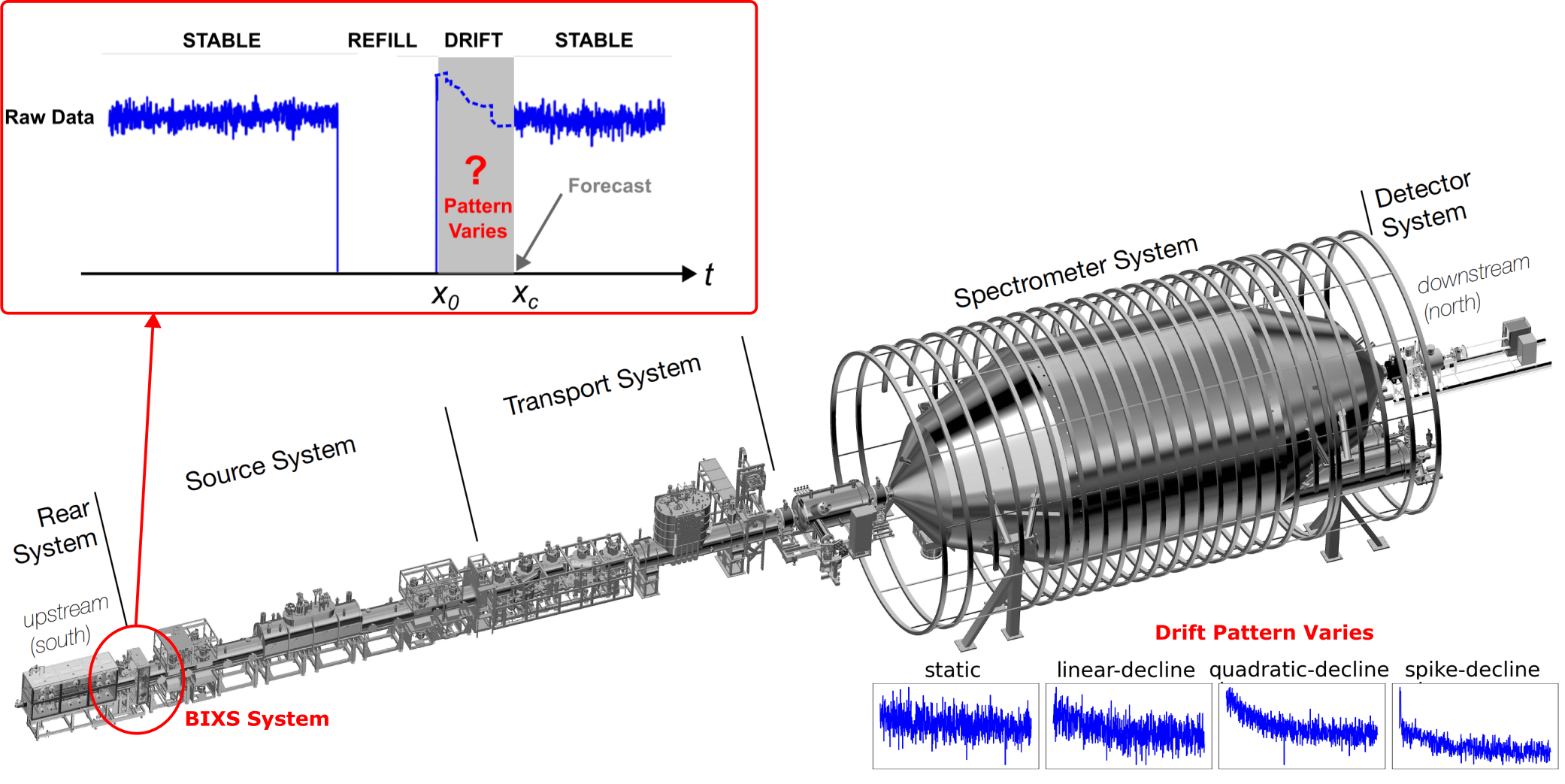}
  \caption{The key challenge is predicting the stable time, $X_c$, of the tritium source. The raw data shows a stable activity level before the refill-event, followed by a drift phase before eventually stabilizing again. Given the unpredictable nature of these drift patterns, the goal is to develop a reliable method to predict when stability will be regained.}
  \label{fig:motivation}
\end{figure*}

Maintaining a highly stable gas quantity inside the WGTS is crucial for accurately determining the neutrino mass, as uncertainties in the gas amount introduce significant systematic errors in the measurement. To monitor source stability on short time scales ($< \mathcal{O}(10)$,h), a variety of sensors have been installed throughout the experiment~\cite{aker2021design}. The $\upbeta$-Induced X-Ray Spectroscopy (BIXS) system continuously monitors the source activity in real time~\cite{Roellig2015bixs}, and this information is used in the data selection process to filter out periods affected by instabilities. However, traditional approaches for detecting or forecasting drift often rely on fixed thresholds or simple predictive models, which are inadequate for capturing the complex, transient patterns observed in gaseous tritium measurements.

Our study focuses on operational forecasting: identifying when the Tritium source reaches a stable state. While the underlying causes of activity drift are well understood within the KATRIN collaboration~\cite{aker2021design}, the Tritium loop is a complex, multi-stage system with hundreds of meters of piping and multiple buffer vessels. The system's behavior resembles that of a weather system, where slow causal changes can propagate through delayed and nonlinear pathways. The sensor signal we analyze represents the final product of this entire chain of interactions. This complexity forms the core motivation of our work: to forecast the point at which the Tritium source stabilizes, despite sparse data and intricate system dynamics.

Traditional estimation methods or simple statistical heuristics are commonly used to determine stabilization, but they are sensitive to subtle shifts in the activity gradient. Even a minor misjudgment in identifying the equilibrium point can lead to premature or delayed restarts—resulting in hours of idle time. In the context of a large-scale experiment with strict scheduling and high operational costs, such delays translate into significant inefficiencies. This underscores the need for more robust and data-driven approaches, such as temporal machine learning models, that can reliably forecast stability under noisy and dynamic conditions.

To better predict the complex transient behavior of the gaseous tritium source, we apply temporal machine learning models to data collected with BIXS from multiple KATRIN Neutrino Mass (KNM) measurement campaigns. These models are well-suited for time-series forecasting, as they can capture intricate, time-dependent patterns that conventional methods often overlook. Specifically, we aim to predict when the tritium source will reach a stable state (Figure~\ref{fig:motivation}).

Our approach begins with preprocessing KATRIN tritium source activity data—regularization, smoothing, and scaling—to structure it into a suitable time-series format. We evaluate a range of forecasting models, from classical linear approaches to advanced deep learning architectures, including N-BEATS, Temporal Fusion Transformer (TFT), NHITS, and TSMixer, as well as specialized pretrained large language models like Chronos. These models are assessed for their ability to predict equilibrium onset under noisy conditions, emphasizing key performance metrics such as accuracy, execution time, and reliability across multiple datasets. Furthermore, we investigate their feasibility for real-time deployment in KATRIN’s operational environment.

This study provides three key contributions:
\begin{enumerate}
    \item We perform a comparative evaluation of temporal forecasting models with diverse architectural paradigms—including recurrent (LSTM), attention-based (TFT), MLP-style (TSMixer), and residual block-based designs (N-BEATS, NHITS)—on tritium source data from the KATRIN experiment. Rather than seeking state-of-the-art performance, the goal is to assess which architectural features offer robustness and interpretability for scientific forecasting under drift.

    \item We introduce a preprocessing pipeline that integrates Savitzky-Golay filtering with piecewise linear segmentation to detect regime transitions in noisy experimental data. We study how varying the smoothing parameters affects forecast stability and alignment across architectures.

    \item We assess the feasibility of real-time application by measuring each model’s training and inference time, forecast variability, and trial-level reliability—establishing architectural trade-offs relevant to deployment in critical monitoring systems.
\end{enumerate}

Beyond its immediate relevance to KATRIN, this work demonstrates the potential of temporal learning models for automating complex time-sensitive predictions in large-scale scientific experiments.


\section{Related Work}

Time-series forecasting has undergone significant advancements, evolving from traditional statistical models to machine learning and, more recently, deep learning architectures~\cite{miller2024survey,chen2023long}. Classical methods such as ARIMA and Exponential Smoothing performed well for simple, stationary data but struggled with non-linearity, complex patterns, and multivariate dependencies~\cite{de2018forecasting}. These limitations led to the development of machine learning models such as Random Forests~\cite{breiman2001random} and Gradient Boosting Machines~\cite{natekin2013gradient}, which improved predictive performance by capturing non-linear relationships and incorporating external covariates. However, their inability to model temporal dependencies effectively limited their utility for sequence-based tasks.

The advent of deep learning transformed time-series forecasting, particularly with Recurrent Neural Networks (RNNs)~\cite{williams1989learning} and Long Short-Term Memory (LSTM) networks~\cite{hochreiter1997long,siami2018forecasting}, which excelled in capturing long-term dependencies. While LSTMs improved sequential modeling, they faced scalability challenges for long sequences and were computationally expensive, prompting further advancements.

To overcome these limitations, non-recurrent neural architectures such as N-BEATS (Neural Basis Expansion Analysis for Time Series)~\cite{oreshkin2019n} and NHITS (Neural Hierarchical Interpolation for Time Series)~\cite{challu2023nhits} were introduced. N-BEATS used fully connected residual blocks to model trends and seasonality effectively, while NHITS extended this approach with hierarchical interpolation for handling long-horizon forecasting. In parallel, Transformer-based models like the Temporal Fusion Transformer (TFT)~\cite{lim2021temporal} leveraged attention mechanisms to selectively focus on key input features, making them particularly suited for multi-horizon forecasting with time-varying covariates. Despite their success, these models introduced new challenges: N-BEATS lacked support for covariates, while Transformer-based models were computationally expensive.

Motivated by the need for more efficient architectures, researchers explored alternative approaches such as NLinear and DLinear~\cite{zeng2023transformers}, which simplified multivariate forecasting by directly mapping inputs to outputs, reducing computational overhead while maintaining competitive accuracy. More recently, TSMixer~\cite{chen2023tsmixer} introduced multilayer perceptron (MLP)-based mixers to capture both temporal and cross-variate dependencies, offering strong performance without reliance on recurrent or attention-based structures.

In 2024, LLM-based time-series forecasting emerged as a dominant paradigm, leveraging the flexibility of pretrained large language models (LLMs). Methods such as PromptCast~\cite{xue2023promptcast} and LLMTime~\cite{gruver2024large} enabled zero-shot forecasting by representing numerical time-series data as text, while fine-tuned models like GPT4TS~\cite{zhou2023one} and Time-LLM~\cite{jin2023time} improved accuracy using domain-specific embeddings. Chronos~\cite{ansari2024chronos} departed from this approach by training a dedicated LLM from scratch on a tokenized time-series corpus, prioritizing scalability and generalization.

To comprehensively evaluate forecasting models for equilibrium prediction in the KATRIN experiment, we intentionally select a representative and diverse set of architectures: LSTM (as a baseline recurrent model), N-BEATS and NHITS (non-recurrent neural approaches), TFT and TSMixer (attention- and MLP-based methods), NLinear and DLinear (lightweight linear models for efficient multivariate forecasting), and Chronos (representing the LLM-based forecasting family). This selection spans major modeling paradigms and is sufficient to reveal architectural trade-offs in our scientific context. To support future benchmarking, we prepared a reproducible and modular workflow, and our code is available on GitHub (\url{https://github.com/nicolaisi/tritium_source_forecasting.git}). Additional models could be integrated, and we welcome such extensions in follow-up studies.
\section{Methodology}

Our approach aims to predict when the instability in the tritium source will return to a stable state by predicting future data points and reconstructing the stabilization curve using BIXS data. Specifically, we detect the onset of what we define as \emph{activity drift} (Figure~\ref{fig:motivation}), which refers to the spike in the value followed by a gradual stabilization after the tritium source is refilled. We introduce a one-hour delay before generating a long-term forecast, projecting the system’s behavior up to thirteen hours into the future.

\subsection{Activity Drift Detection}

To detect the onset of activity drift, we identify periods of extremely low data values, typically near zero, followed by a sharp increase when the tritium source is refilled. The activity then gradually decreases toward equilibrium. Our approach relies on detecting significant changes between consecutive data points using a simple threshold-based method (Algorithm~\ref{alg:activity_drift}).

\begin{algorithm}[!htb]
    \DontPrintSemicolon
    \textbf{Input:} Time-series data $D$ with timestamps $t$ and values $y_t$\;
    \textbf{Output:} Detected jumps $J$ and descending points $P_{\text{desc}}$\;
    $\Delta y \gets \text{difference of } y_t$\;
    $T_{\text{jump}} \gets 500.0$\;
    $T_{\text{value}} \gets 2500.0$\;
    $is\_jump[t] \gets (\Delta y_t > T_{\text{jump}}) \land (y_t > T_{\text{value}})$\;
    $J \gets \{t : is\_jump[t]\}$\;
    \SetKwFunction{FindDescendingPoints}{FindDescendingPoints}
    \SetKwProg{Fn}{Function}{:}{}
    \Fn{\FindDescendingPoints{$D$, $J$}}{
        $P_{\text{desc}} \gets \{\}$ \tcc*[r]{Initialize descending points}
        \For{$t_{\text{jump}} \in J$}{
            $start\_pos \gets \text{index of } t_{\text{jump}}$\;
            \For{$i \gets start\_pos+1$ \KwTo $\text{len}(D)$}{
                \If{$y_i < y_{i-1}$}{
                    $P_{\text{desc}} \gets P_{\text{desc}} \cup \{t_i\}$\;
                    \textbf{break}\;
                }
            }
        }
        \textbf{return} $P_{\text{desc}}$\;
    }
    $P_{\text{desc}} \gets \FindDescendingPoints(D, J)$\;
    $is\_descending[t] \gets t \in P_{\text{desc}}$\;
    $D_{\text{desc}} \gets \{t : is\_descending[t]\}$\;
    \caption{Activity Drift Start Point Identification}
    \label{alg:activity_drift}
\end{algorithm}

\subsection{Fixed-length Prediction}
We approach the fixed-length forecasting problem in discrete time, extending the univariate case~\cite{oreshkin2019n} to multivariate scenarios. For univariate forecasting, given a forecast horizon of length $H$ and an observed history $[y_1, y_2, ..., y_T] \in {\rm I\!R}^T$, the task is to predict the vector of future values $y \in {\rm I\!R}^H=[y_{T+1}, y_{T+2}, ..., y_{T+H}]$. In this context, we use a lookback window of length $t \le H$, starting with the first observed value $y_t$ as the model input, denoted $x \in {\rm I\!R}^t=[y_1, y_2, ..., y_t]$, and we denote the forecast as $\hat{y}$.

For the multivariate case, where multiple independent time series are forecast simultaneously, each series has its own observed history $[y_1^i, y_2^i, ..., y_T^i]$ and forecast horizon $[y_{t+1}^i, y_{t+2}^i, ..., y_{t+H}^i]$ for $i=1, 2, ..., N$. Here, the lookback window $x^i \in {\rm I\!R}^t$ ensures that each series’ unique temporal pattern is captured, while maintaining the independence of their forecasts. Typically, the same lookback window length $t$ is applied to all time series, with each series using its own first $t$ observed data points as input.

In the case of the tritium source, the activity typically remains stable, but a sharp drop in value indicates depletion. To model this, we track data trends for one hour following the onset of activity drift. This one-hour period forms our lookback window, where $x \in {\rm I\!R}^t$ corresponds to 36 data points (each sampled at 100-second intervals). The model uses these 36 data points to learn the dynamics immediately after the activity drift begins, providing critical context for the system’s stabilization.


From this input, our temporal models forecast $H=460$ future data points, corresponding to a 12.7-hour forecast horizon. This horizon is chosen to cover the entire stabilization process until equilibrium is reached. Each forecasted point represents a 100-second interval, forming the output vector $\hat{y} \in {\rm I\!R}^H=[\hat{y}_{t+1}, \hat{y}_{t+2}, ..., \hat{y}_{t+H}]$.

This method can be applied at the onset of activity drift or deployed as part of a real-time monitoring system, offering valuable insights into the stabilization process and enabling timely interventions or adjustments.

\begin{figure}[t]
  \centering
      \includegraphics[width=0.45\textwidth]{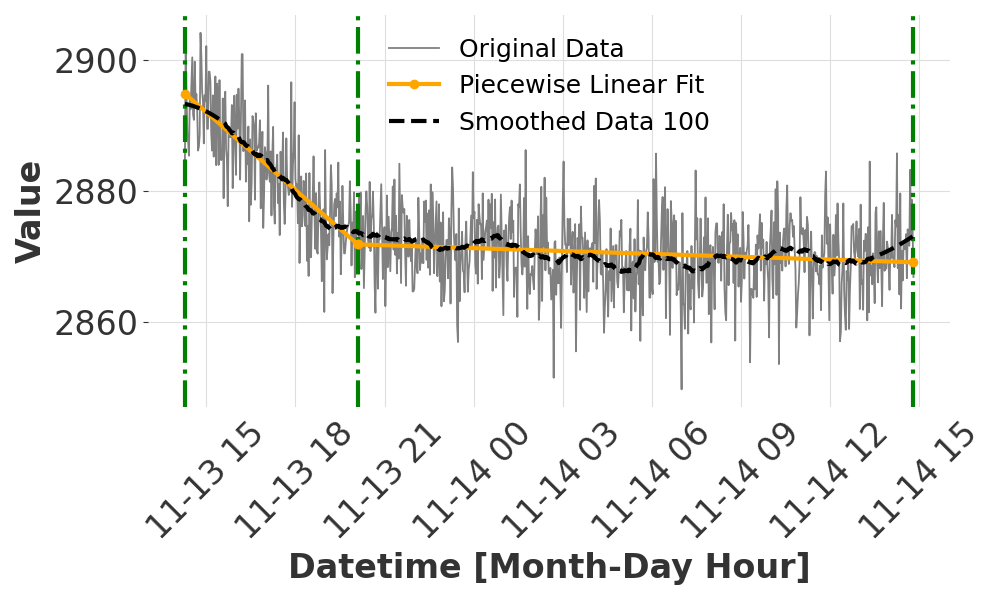}
  \caption{Savitzky-Golay filtering and two-segment piecewise linear fitting applied to raw data.}
  \label{fig:sg_line}
\end{figure}

\begin{figure}[htb]
  \centering
      \includegraphics[width=0.45\textwidth]{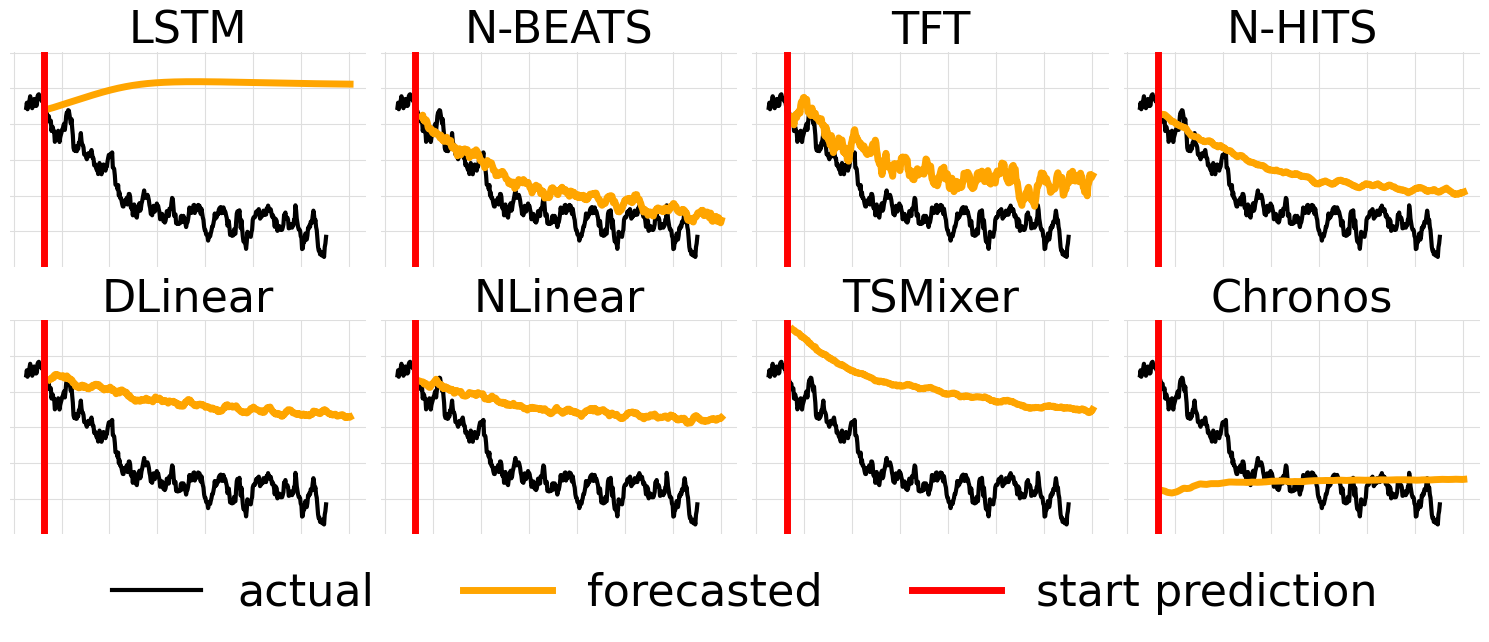}
  \caption{Overview of the predictions for Dataset 26 generated using state-of-the-art models.}
  \label{fig:overview_forecasting_26}
\end{figure}

\begin{table*}[ht]
\centering
\caption{Optimal Hyperparameters and Search Ranges for Each Model}
\begin{tabularx}{\textwidth}{lclX}
\toprule
\textbf{Model} & \textbf{Loss $L(\theta)$} & \textbf{Trials} & \textbf{Hyperparameters (Optimal [Min, Max])} \\
\midrule
LSTM & 2.17 & 50 & n\_epochs: 38 [10, 100], batch: 64 [16, 64], dropout: 0.1072 [0.0, 0.5] \\
N-BEATS & 1.47 & 50 & n\_epochs: 70 [10, 100], batch: 64 [16, 64], dropout: 0.0187 [0.0, 0.5], blocks: 4 [1, 10], hidden: 160 [8, 256] \\
TFT & 1.31 & 50 & n\_epochs: 70 [10, 90], batch: 32 [16, 64], dropout: 0.1153 [0.0, 0.5], LSTM layers: 2 [1, 4], hidden: 16 [8, 128] \\
NHITS & 1.50 & 50 & n\_epochs: 23 [10, 100], batch: 32 [16, 64], dropout: 0.2038 [0.0, 0.5], downsample: 1 [1, 4], layers: 3 [1, 4], blocks: 4 [1, 5], stacks: 5 [1, 5] \\
DLinear & 1.36 & 50 & n\_epochs: 88 [10, 100], batch: 16 [16, 64], kernel: 2 [1, 5] \\
NLinear & 1.93 & 50 & n\_epochs: 30 [10, 100], batch: 16 [16, 64], dropout: 0.2038 [0.0, 0.5] \\
TSMixer & 1.19 & 50 & n\_epochs: 90 [10, 100], batch: 16 [16, 64], dropout: 0.4620 [0.0, 0.5], blocks: 2 [1, 10], hidden: 8 [8, 128] \\
\bottomrule
\end{tabularx}
\label{tab:optimal_hyper}
\end{table*}

\subsection{Incremental Lookback Window}
\label{subsec:lookback_window}

Since our solution integrates with a real-time data monitoring system, we analyze how prediction evolves as historical context gradually increases. As the system nears equilibrium, we expect predictions to improve due to the model's ability to leverage a longer history of observed trends.

The model starts with a minimal lookback window $t_0$ (e.g., a single data point). With each iteration, the lookback window grows incrementally, incorporating one additional data point per step (i.e., $t_0, t_0+1, t_0+2, ...$). At timestep $k$, the lookback window contains $k$ historical values, $x^i = [y_1^i, y_2^i, ..., y_k^i] \in {\rm I\!R}^k$, where $k$ follows the recurrence relation: $k_{n+1} = k_n + 1$.

Here, $n$ denotes the iteration step, and $k_n$ represents the lookback window length at step $n$. This incremental approach allows the model to progressively incorporate more historical context, improving its ability to predict stabilization timing and refine long-horizon forecasts.

\subsection{Determination of First Equilibrium Point}

After forecasting a set of data, we determine when the tritium source stabilizes. To achieve this, we first apply the Savitzky-Golay filter~\cite{savitzky1964smoothing, schafer2011savitzky} to smooth the forecasted data while preserving its structure. Next, we fit the smoothed data with a two-segment piecewise linear model~\cite{jekel2019pwlf}. Importantly, the smoothing algorithm is used only to aid in identifying the equilibrium point and does not influence the forecast itself.

Given a set of data points $y_1, y_2, ..., y_n$, the Savitzky-Golay filter smooths each point $y_i$ by replacing it with a weighted sum of neighboring values within a fixed window. Mathematically: $\hat{y_i} = \sum_{j=-k}^{k} c_j y_{i+j}$, where $k$ is half of the window size $W$, meaning $W = 2k + 1$. The weights $c_j$ are convolution coefficients that depend on the polynomial order $p$ and the window length $W$. The values $y_{i+j}$ represent neighboring data points centered around $y_i$~\cite{savitzky1964smoothing}. The equilibrium point is identified as the start of the flat segment. Mathematically, the piecewise function is given by: \begin{equation*}
    y_{i} =
    \begin{cases}
      m_1 \hat{y_i} + b_1, & \text{if $x_i \leq x_c$} \\
      m_2 \hat{y_i} + b_2, & \text{if $x_i > x_c$}
    \end{cases}   
\end{equation*}

\noindent where $\hat{y_i}$ is the smoothed value computed by the Savitzky-Golay filter, $m_1, b_1$ are the slope and intercept of the first segment (decline phase), $m_2, b_2$ are the slope and intercept of the second segment (stabilization phase), and $x_c$ represents the breakpoint between the two segments. Figure~\ref{fig:sg_line} illustrates the smoothed data overlaid with the two-segment piecewise linear fit.

\subsection{Loss Function}

Training the models involves minimizing the total breakpoint prediction error across all validation datasets. We denote the predicted breakpoint from the two-segment piecewise linear regression as $x_c$ and the true breakpoint as $x_{\text{actual}}$. Our objective is to minimize their absolute difference. To generalize across multiple test datasets ($\Omega$), we define the total loss as $ \mathcal{L}(\theta) = \frac{1}{|\Omega |} \sum_{x \in \Omega} |x_c(\theta) - x_{\text{actual}}| $, where $\Omega$ represents the set of all test datasets and $x_c(\theta)$ is the predicted breakpoint for dataset $x$ given model parameters $\theta$. The optimal hyperparameters $\theta^*$ are obtained by minimizing the loss across all datasets: $\theta^* = \arg\min_{\theta} \mathcal{L}(\theta)$. To efficiently search for $\theta^*$, we employ the Optuna framework~\cite{akiba2019optuna}, a Bayesian optimization approach leveraging Tree-structured Parzen Estimators (TPE). Optuna's adaptive sampling strategy allows efficient hyperparameter tuning, reducing the computational cost of exhaustive grid search. Table~\ref{tab:optimal_hyper} provides full details on the hyperparameter optimization process and parameters. The selected models are optimized by minimizing the forecasted stability time and the actual stability time. We run the optimization on a Jupyter notebook connected to an NVIDIA A100 GPU. For the Chronos model, we select the best pre-trained model (chronos-large) from the others (chronos-small, chronos-base, chronos-bolt-base) and use it in our evaluation. 



\begin{figure*}[t!]
  \centering
  \subfloat[MAE]{%
    \includegraphics[width=0.32\textwidth]{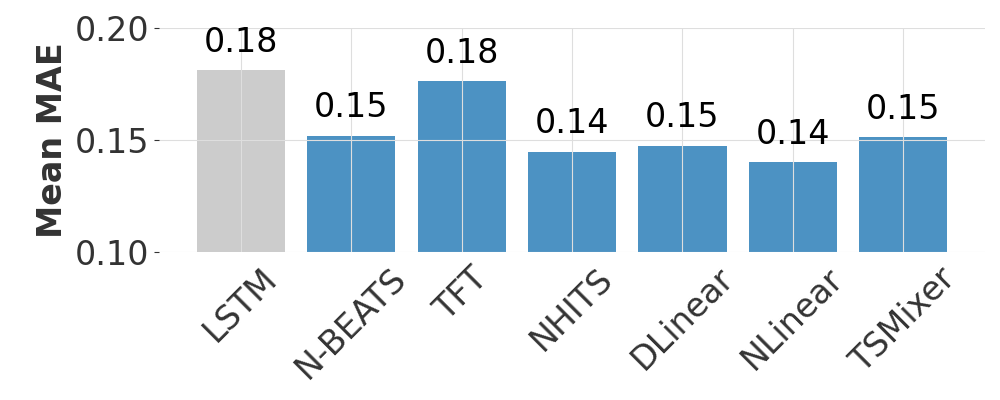}%
    \label{fig:subfig2}
  }\hfill
  \subfloat[sMAPE]{%
    \includegraphics[width=0.32\textwidth]{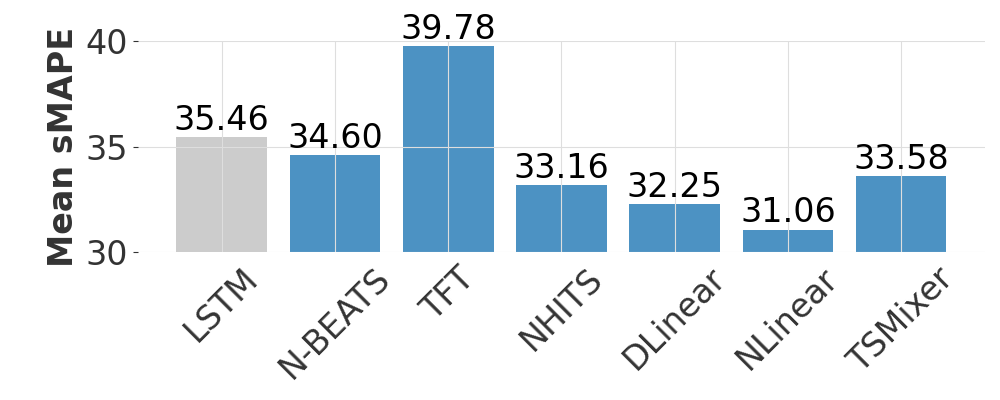}%
    \label{fig:subfig3}
  }\hfill
  \subfloat[DTW]{%
    \includegraphics[width=0.32\textwidth]{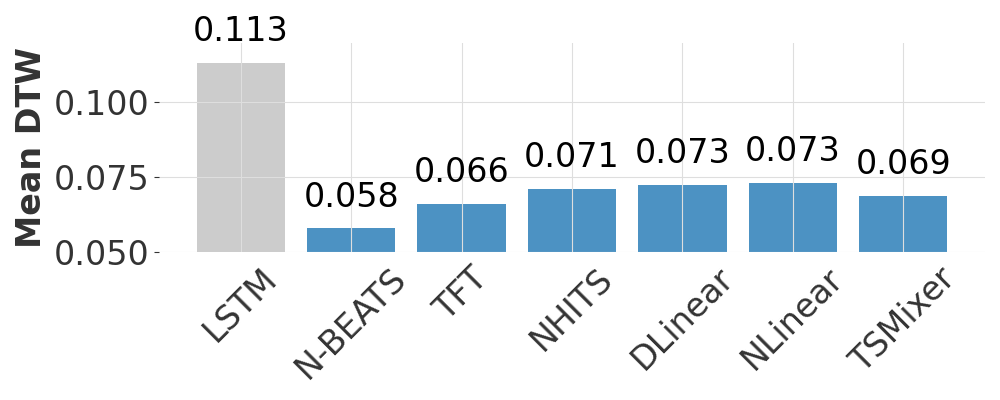}%
    \label{fig:subfig4}
  }
  \caption{Comparison metrics of the forecasted data against the real data. Models excluded from consideration in earlier evaluations are indicated by the grey label.}
  \label{fig:metrics}
\end{figure*}

\section{Evaluation and Discussion}

Our analysis used BIXS measurement data from May 11, 2020, to September 19, 2024, which comprises 33 instability events. Of these, 25 datasets were allocated for training and 8 for validation. It is important to note that the raw noisy data was used for training without applying any smoothing algorithm. In this work, we consistently refer to the validation datasets as Datasets 26–33.

We evaluated model performance on an NVIDIA A100 GPU within a Jupyter notebook. We implemented most models using the Darts~\cite{JMLR:v23:21-1177} framework, while Chronos relied on GluonTS~\cite{JMLR:v21:19-820}. We preprocessed the irregular raw datasets by regularizing them to 100-second intervals before scaling them for prediction. For reproducibility, the datasets and detailed analysis notes is available on GitHub~\footnote{\url{https://github.com/nicolaisi/tritium_source_forecasting.git}}.

\subsection{Forecast Accuracy and Model Behavior}

Figure~\ref{fig:overview_forecasting_26} shows how well each model’s forecast matches the real data. The goal is to use model predictions to estimate the equilibrium’s starting point based on forecast accuracy. At the beginning of the activity drift, we allowed an initial waiting period of one hour; this first hour of data was used as input for the forecasting model’s lookback window. The red vertical line indicates the time at which 460 future data points were predicted. 

Only the LSTM and Chronos models failed to capture the downward trend shown in the plots. Rather than forecasting a downward trend, the Chronos LLM model predicts a nearly stable trajectory, allowing for real-time estimation of proximity to equilibrium. However, since Chronos predicts a behavior that is not relevant to the focus of this study, it is excluded from further evaluations and will not be included in our analysis henceforth. LSTM was excluded from consideration due to its inability to capture the downward trend. Excluded models will be greyed out to indicate their removal from the selection process.

\begin{figure*}[htb]
  \centering
      \includegraphics[width=0.90\textwidth]{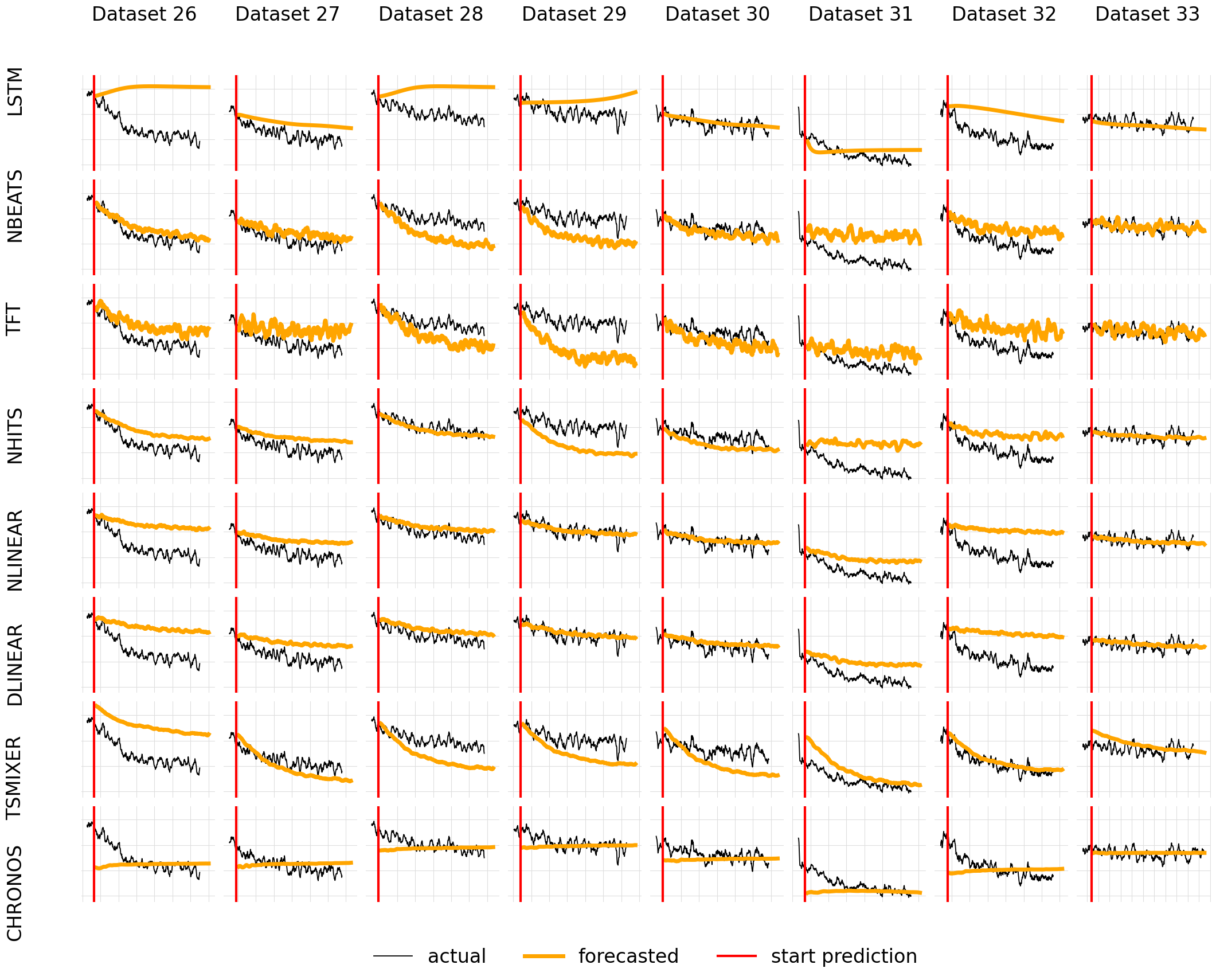}
  \caption{Overview of the forecasted data based on state-of-the-art models.}
  \label{fig:model_forecasting}
\end{figure*}

Notably, the N-BEATS and TFT models introduce noise that closely resembles the underlying trend. Simpler models, such as DLinear and NLinear, exhibit strong linearity. Among all the models, N-BEATS, TFT, NHITS, and TSMixer produced forecasts accurate enough for reliable equilibrium point predictions. Figure~\ref{fig:model_forecasting} provides a visual overview of each model’s forecasting curve across the validation datasets.

Figure~\ref{fig:metrics} presents the metrics used to evaluate each model’s prediction performance. To assess forecast accuracy, we compute the Mean Absolute Error (MAE) and Symmetric Mean Absolute Percentage Error (sMAPE). Higher MAE and sMAPE values in N-BEATS and TFT indicate noisier forecasts. However, this noise does not prevent accurate identification of the equilibrium start point, as our method prioritizes the overall curve shape over pointwise precision. This ensures that the breakpoint between the two segments is reliably detected.  
While MAE and sMAPE measure forecast deviation from actual values, Dynamic Time Warping (DTW) quantifies curve shape similarity, which is more relevant to our goal. Figure~\ref{fig:subfig4} shows that N-BEATS achieves the lowest DTW, highlighting its ability to accurately capture the overall curve shape.  

\begin{figure}[htb]
  \centering
      \includegraphics[width=0.45\textwidth]{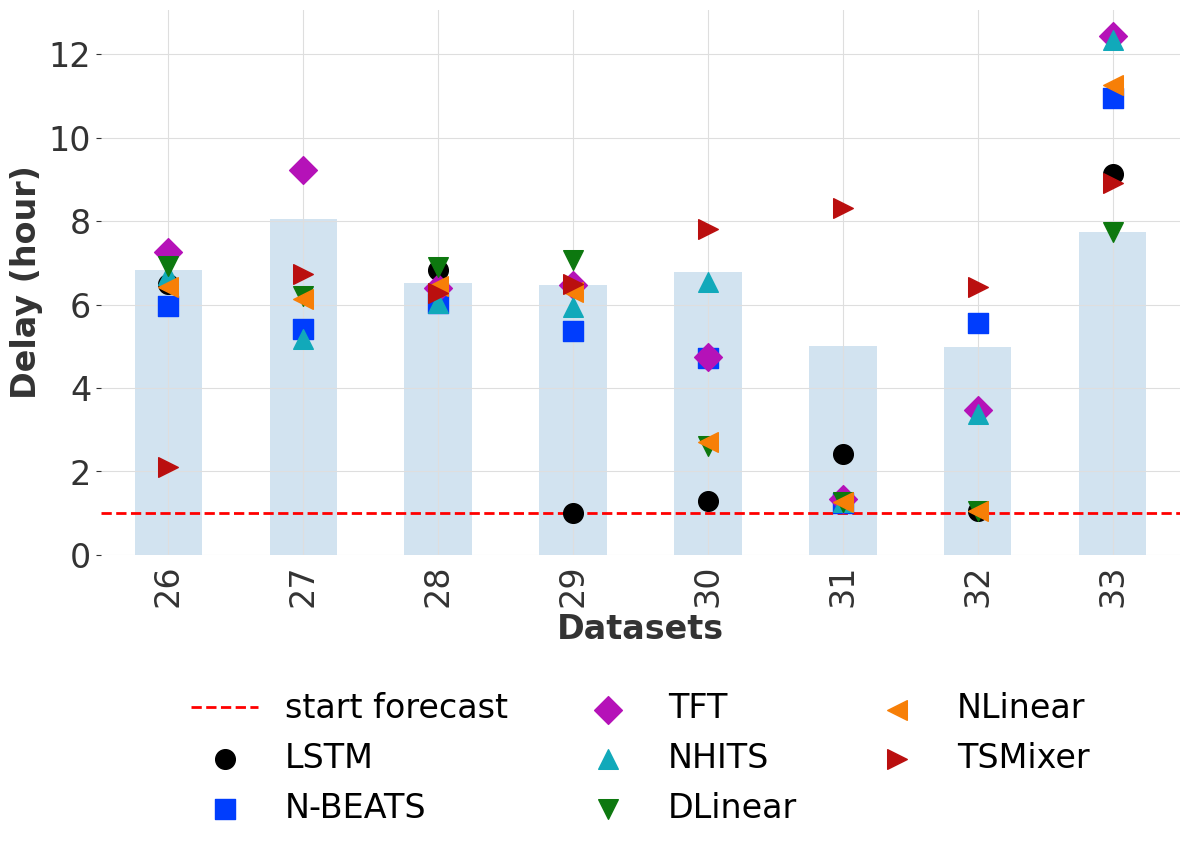}
  \caption{The bar chart shows the delay in stability predictions by the models, represented by colored symbols, alongside the actual stabilization time, represented by the blue bar chart, for each dataset. Forecasting begins one hour after the activity drift is detected, as marked by the red dashed line. The predictions reflect the average results across 50 iterations.}
  \label{fig:delay_predict}
\end{figure}

\subsection{Predicting the Start of Equilibrium}

This study investigates the ability of temporal learning models to predict the onset of the equilibrium line. To assess model performance, we compare each model’s predicted start time to the actual start of equilibrium.

Figure~\ref{fig:delay_predict} illustrates the alignment between predicted data points (colored dots) and actual stability times (blue bars). Prediction begins one hour after detecting the activity drift, marked by the red dotted line. Each data point within the bar represents the predicted time to reach a stable state. Accurate timing is crucial, as predictions occurring too early or too late can impact usability. For optimal user experience, forecasted points should closely align with actual stability time to enable a quick resumption of measurements.

\begin{figure*}[t!]
  \centering
  \subfloat[]{%
    \includegraphics[width=0.32\textwidth]{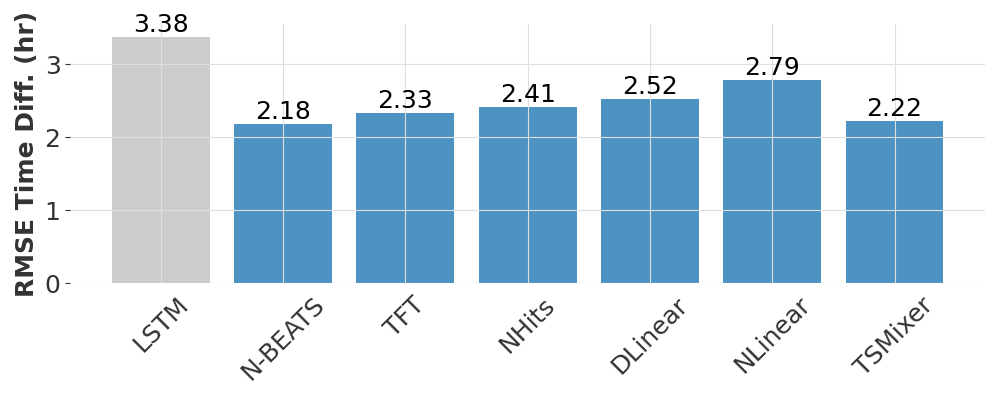}%
    \label{fig:scatter_time}
  }\hfill
  \subfloat[]{%
    \includegraphics[width=0.32\textwidth]{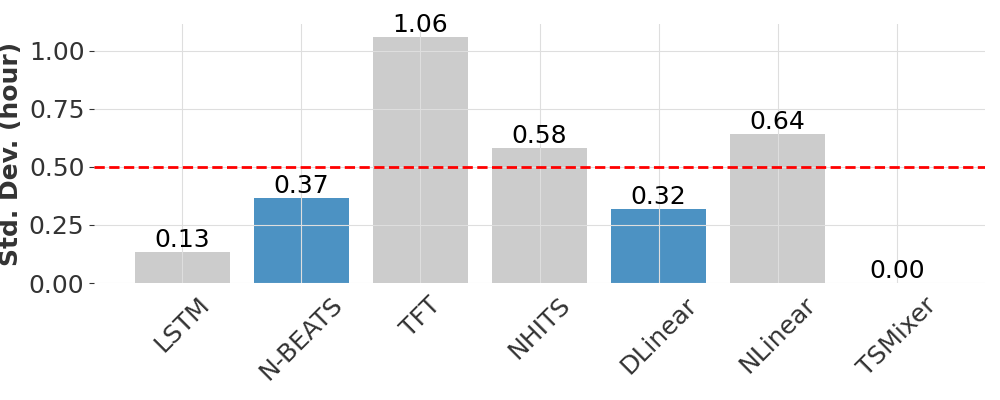}%
    \label{fig:mode_stdev}
  }\hfill
  \subfloat[]{%
    \includegraphics[width=0.32\textwidth]{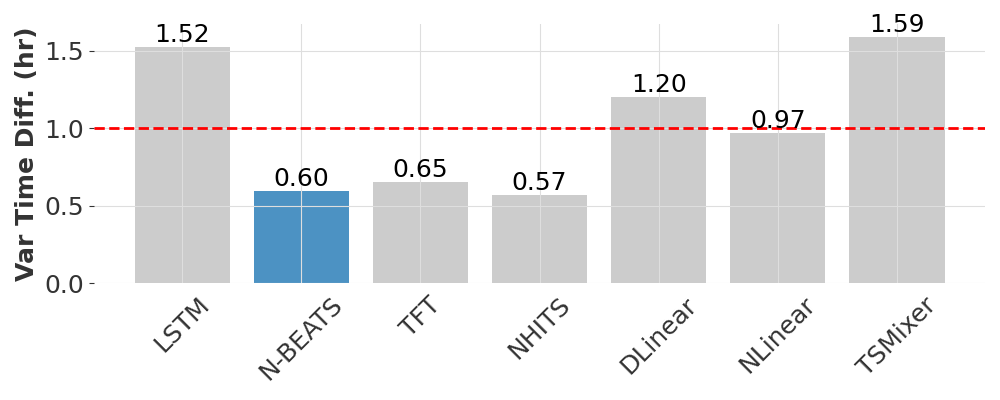}%
    \label{fig:time_diff_moving_step}
  }
  \caption{(a) The plot displays the RMSE between the predicted time and the actual stability time, with models excluded from earlier evaluations highlighted in grey. (b) The bar chart shows the standard deviation of each model after 50 iterations. (c) Variance of the time difference for the moving prediction.}
  \label{fig:metrics2}
\end{figure*}

The analysis focuses on how well predicted points align with actual stable times. The plot also highlights the variability in model performance across datasets, indicating no universally superior model. To better understand the differences between predicted and actual times, Figure~\ref{fig:scatter_time} presents the Root Mean Square Error (RMSE) time difference for each model across all datasets. Among the models, N-BEATS performs slightly better, achieving the lowest RMSE time difference.

\subsection{Prediction Reliability}

Challu et al.~\cite{challu2023nhits} observed that long-term time series forecasting accuracy tends to decrease as the prediction horizon increases (measured by MAE). Given our prediction horizon of 460 data points, a higher MAE is therefore expected. Consequently, our primary goal is to assess whether the predictions maintain stability over repeated iterations or exhibit significant variability. 


To evaluate model stability, we run 50 iterations per model and calculate the standard deviation of their predictions. Figure~\ref{fig:mode_stdev} highlights the stability of TSMixer, reflected in its low standard deviation across iterations. In contrast, TFT’s higher standard deviation stems from its sensitivity to noisy features. Models with a standard deviation exceeding 30 minutes, such as NHITS, were excluded from further consideration.

\subsection{Incremental Lookback Window}

Section~\ref{subsec:lookback_window} details the incremental lookback window used to assess the real-time performance of our predictive models. In a real-time system, predictions are updated regularly, so it is valuable to evaluate how predictions evolve from the start of activity drift to its stabilization. Theoretically, our prediction should improve as it approaches the actual stable time, thanks to a larger historical context.

\begin{figure}[htb] \centering \includegraphics[width=0.45\textwidth]{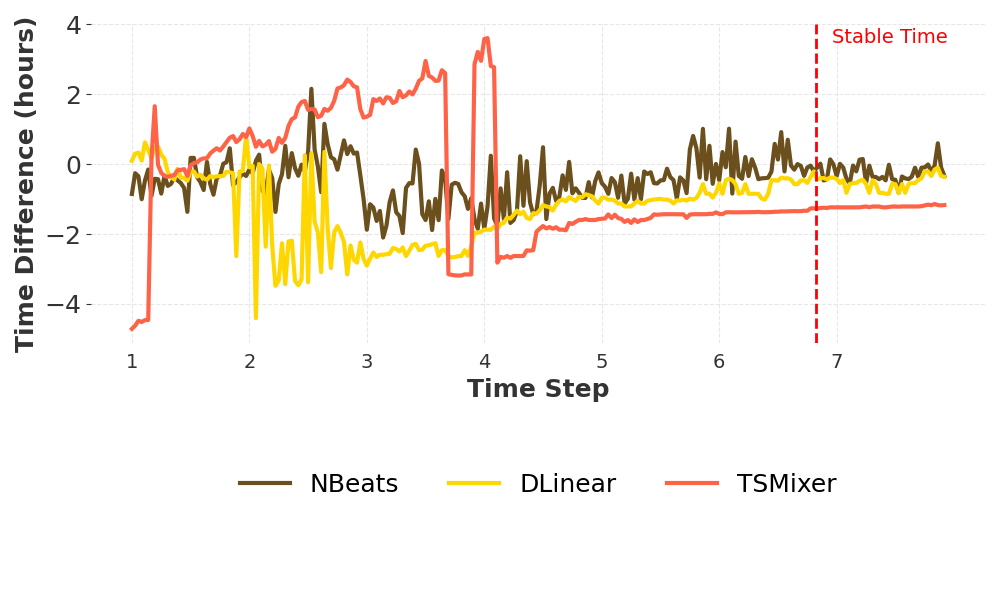} \caption{The prediction performance of the final three models (N-BEATS, DLinear, TSMixer) was evaluated on Dataset 26, with predictions generated at each step, corresponding to a 100-second interval. The predictions represent the average results over 50 iterations.} \label{fig:moving_step} \end{figure}

We use Dataset 26 and incrementally increase the lookback time in steps of 100 seconds. The evolution of the predictions is depicted in Figure~\ref{fig:moving_step}. We start with a one-hour lookback time, where the values at the first step correspond to those in the first bar of Figure~\ref{fig:delay_predict}.


The prediction performance of the models is shown in Figure~\ref{fig:moving_step} as the prediction point is moved one step at a time. The predictions eventually converge toward the stable time (the start of equilibrium). However, TSMixer’s one-hour offset reduces its reliability, despite its good repeatability score. The variance of the time difference between predicted and actual values for each model at every step is shown in Figure~\ref{fig:time_diff_moving_step}.Among the remaining models—N-BEATS, DLinear, and TSMixer—an ideal scenario would feature minimal variance. The choice of a 30-minute threshold was based on empirical assessment, as a variance of one hour is considered suboptimal. This threshold represents a balanced trade-off between stability and sensitivity. Applying this criterion, N-BEATS emerges as the preferred model, as both DLinear and TSMixer demonstrate poor performance in real-time predictions.

\begin{figure*}[htb]
  \centering
  \subfloat[]{%
    \includegraphics[width=0.32\textwidth]{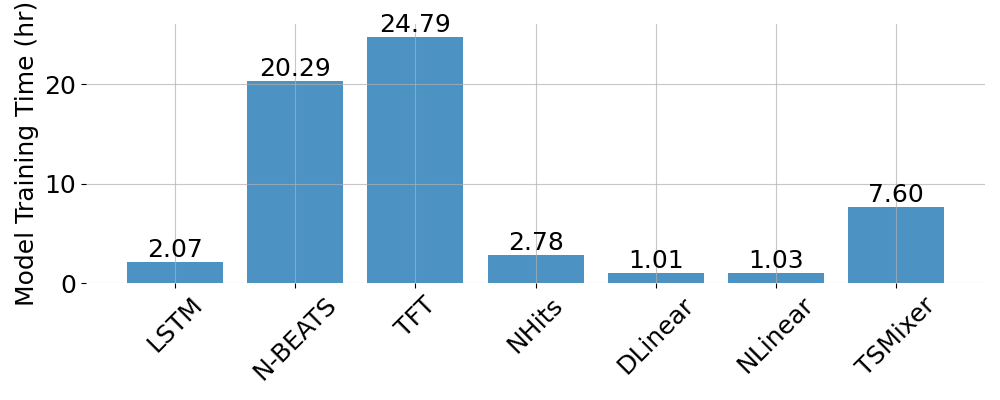}%
    \label{fig:training_time}
  }\hfill
  \subfloat[]{%
    \includegraphics[width=0.32\textwidth]{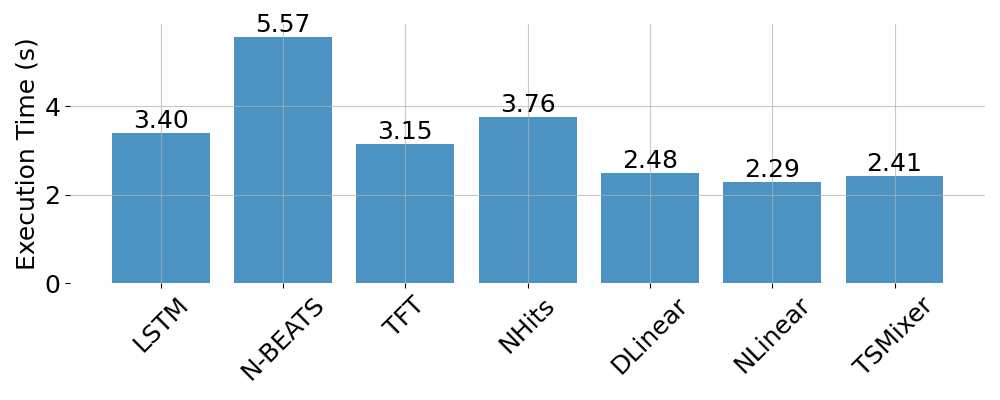}%
    \label{fig:exec_time}
  }\hfill
  \subfloat[]{%
    \includegraphics[width=0.32\textwidth]{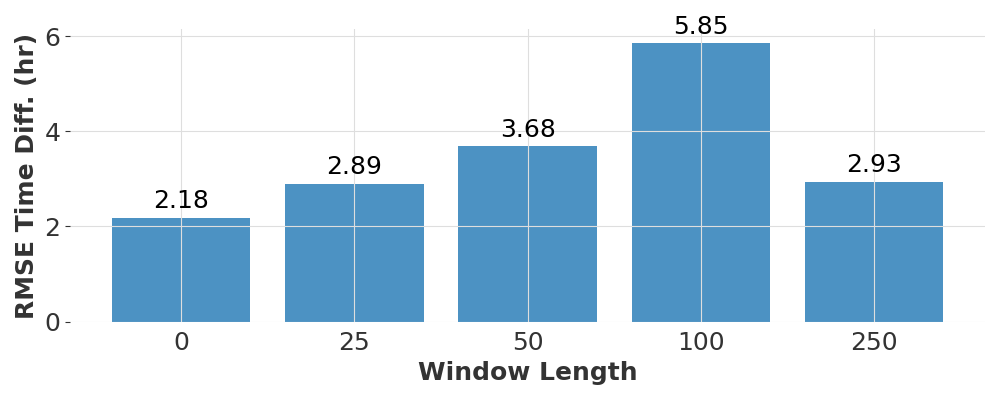}%
    \label{fig:averaging_bar}
  }
  \caption{(a) Training time for each model based on 15 trials (hours). (b) Prediction time per model on the validation datasets (seconds). (c) Time difference between forecasted and actual stability time for N-BEATS variants trained with different smoothing window lengths; the first bar is the model trained on unsmoothed data.}
  \label{fig:processing_time}
\end{figure*}

\subsection{Effect of Smoothing Training Datasets}

\begin{figure}[htb]
  \centering
      \includegraphics[width=0.45\textwidth]{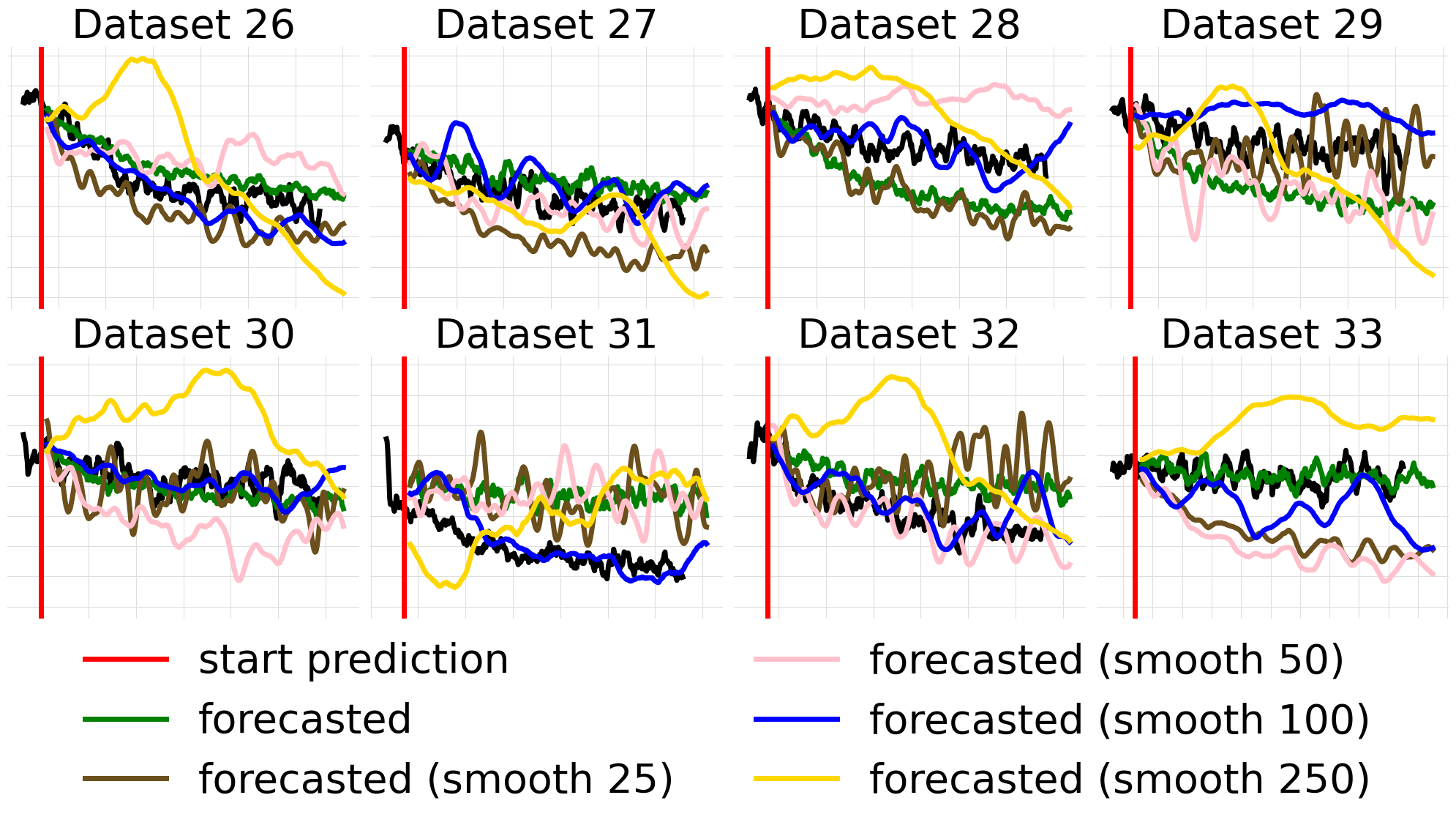}
  \caption{The prediction performance of the N-BEATS model was evaluated across eight test datasets, with each variant trained using a distinct window length.}
  \label{fig:avgeraging_line}
\end{figure}

Noisy training data suggests that smoothing could potentially improve model training. This hypothesis is evaluated by adjusting the window length, where longer lengths result in less noisy datasets. Since we have finalized N-BEATS as the most suitable model for our application, we smooth the training datasets using different window lengths and use them for training. The effect of varying smoothing window lengths on the N-BEATS model’s final result is shown in Figure~\ref{fig:avgeraging_line}. Surprisingly, the best performance came from the unsmoothed training datasets (green line), which contradicts our original hypothesis. According to Figure~\ref{fig:averaging_bar}, the dataset without smoothing exhibits the smallest time difference. Although the RMSE decreases for very large windows, we dismiss this as an improvement since the predicted curves are visibly inaccurate. 


\begin{table*}[htb]
\caption{Events during the KNM15 measurement campaign.}
    \label{tab:post_events}
    \centering
    \begin{tabularx}{0.95\textwidth}{|l|l|X|}
\hline
\textbf{Event} & \textbf{Date/Time} & \textbf{Expert Notes} \\
\hline
E1 & 2024-09-19 11:42:54 & refill from empty source to nominal (gas density)\\
\hline
E2 & 2024-10-01 06:58:00 & refill from empty source to nominal\\
\hline
E3 & 2024-10-10 09:41:39 & refill from empty source to nominal ($\approx$ \SI{1}{\hour} empty state, beforehand $\approx$ 50\% of nominal level)\\
\hline
E4 & 2024-11-14 07:29:21 & increase from $\approx$ 25\% to nominal, with slight $\approx$ 1\% re-increase after \SI{1.5}{\hour}\\
\hline
E5 & 2024-11-21 14:18:27 & refill from nearly empty source to nominal (only \SI{11}{\hour} of nominal level data)\\
\hline
E6 & 2024-11-29 09:10:11 & refill from empty source to nominal ($\approx$ \SI{1}{\hour} empty state, beforehand $\approx$ 50\% of nominal level)\\
\hline
    \end{tabularx}
    
\end{table*}

\subsection{Real-world Deployment}

For real-world implementation, the N-BEATS model was integrated within the BORA framework~\cite{jerome2024bora}, a compact data monitoring system used by the KATRIN experiment. The framework prioritizes rapid prototyping and accessibility, making it ideal for real-time data processing.

The monitoring interface periodically refreshes real-time experimental data, which is continuously written to a Redis database. The BORA framework leverages OpenShift for scalability, reliability, and easy deployment. Containerized applications are supported in this environment, enabling streamlined updates and efficient resource management.

We developed a Python module within the framework to detect activity drift and forecast equilibrium start times. Using predefined inputs, this module forecasts and generates image plots, integrating seamlessly with the framework’s visualization pipeline. Since the temporal learning models are trained offline, the forecasting process operates efficiently with minimal computational resources, removing the need for GPU support. Moreover, the framework accommodates other models for comparative analysis, enhancing the adaptability of the monitoring setup. The training and execution times, shown in Figure \ref{fig:training_time} and \ref{fig:exec_time} justified the offline training approach, as model training takes hours, while inference with a pre-trained model requires only seconds.

The value of this work is demonstrated both through its evaluation metrics and its real-world scientific impact. This framework simplifies a previously complex and time-consuming task, aiding scientists in predicting when the tritium source will reach equilibrium. Furthermore, the automation of this method enables more efficient measurement campaigns and sets a precedent for similar applications in other time-series forecasting domains.



\subsection{Post-launch Performance}

In the KATRIN experiment, multiple measurement campaigns are conducted, with each KATRIN Neutrino Mass (KNM) campaign comprising several hundred $\beta$-spectrum scans. The scientific findings from KNM1-5 have been published~\cite{aker2024direct}, while this study utilizes data from KNM3-14. This work was initiated during KNM13 and implemented in KNM14. Consequently, this post-launch performance assessment evaluates the findings from KNM15, which includes a total of 6 events (Table~\ref{tab:post_events}). Our post-launch evaluation consists of two key aspects: improving activity drift detection and assessing real-time prediction performance.

\subsubsection{Refining Activity Drift Detection}

\begin{figure}[htb] \centering \includegraphics[width=0.48\textwidth]{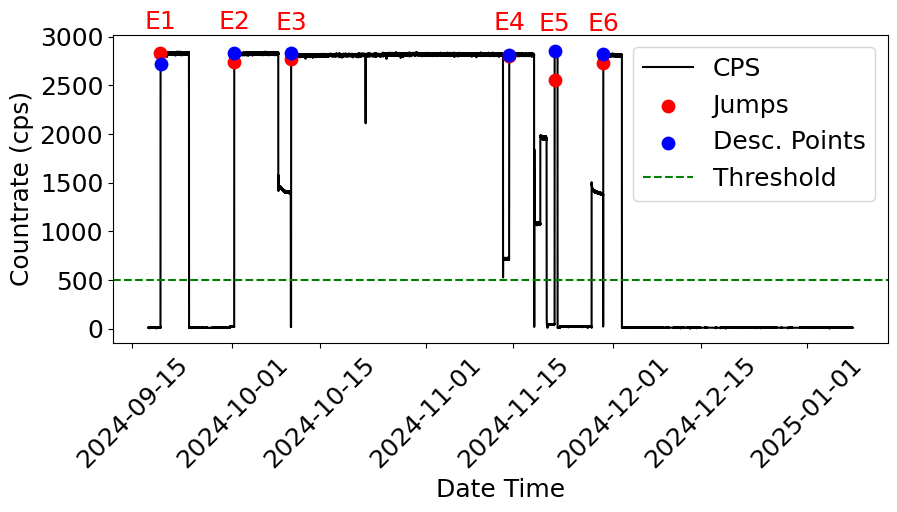} \caption{Time-series profile of the tritium source count rate (counts per second) during KNM15. Red points indicate detected jumps, while blue points mark the start of activity drift.} \label{fig:knm_15} \end{figure}

A sudden and brief dip in the count rate led to an incorrect detection of the activity drift start point when using our simple threshold-based measurement. In the E1 event (Figure~\ref{fig:knm_15}), the peak value does not align with the blue-marked descending point. Since Algorithm~\ref{alg:activity_drift} identifies the first downward point, it failed to capture the actual start of activity drift.

To address this, we adjusted the algorithm: after the jump point, it now considers the next 10 descending points and selects the first one where the absolute delta y is below 25 ($|\Delta y| < 25$). This refinement helps eliminate false detections caused by sharp drops.

\begin{table}[htb]
    \caption{Forecasting performance of the N-BEATS model across the six events in the KNM15 measurement campaign.}
    \label{tab:post_eval}
    \centering
    \begin{tabularx}{0.45\textwidth}{|l|X|X|X|X|}
\hline
\textbf{Event} & \textbf{$\mathcal{L}_0(\theta)$} & \textbf{DTW} & \textbf{MAE} & \textbf{sMAPE} \\
\hline
E1 & -0.599 & 0.039 & 0.061 & 9.497\\
\hline
E2 & 0.834 & 0.057 & 0.132 & 30.497\\
\hline
E3 & -1.170 & 0.048 & 0.102 & 22.641\\
\hline
E4 & -0.169 & 0.056 & 0.106 & 21.462\\
\hline
E5 & -0.068 & 0.093 & 0.212 & 47.315\\
\hline
E6 & -4.765 & 0.047 & 0.108 & 19.249\\
\hline
    \end{tabularx}
    
\end{table}

\begin{table}[htb]
    \caption{Forecasting performance for Event 6 with different shifted start prediction times. $t_{0}$ represents the initial prediction time, while $t_{n}$ represents the $n$th hour after the start of the prediction.}
    \label{tab:post_eval_six}
    \centering
    \begin{tabularx}{0.45\textwidth}{|l|X|X|X|X|}
\hline
\textbf{Start Time} & \textbf{$\mathcal{L}(\theta)$} & \textbf{DTW} & \textbf{MAE} & \textbf{sMAPE} \\
\hline
$t_0$ & -4.765 & 0.039 & 0.061 & 9.497\\
\hline
$t_1$ & -2.277 & 0.044 & 0.103 & 19.598\\
\hline
$t_2$ & -0.667 & 0.047 & 0.089 & 17.013\\
\hline
$t_3$ & -0.169 & 0.056 & 0.106 & 21.462\\
\hline
    \end{tabularx}
    
\end{table}

\subsubsection{Real-Time Prediction Performance}

We evaluated the effectiveness of our approach by comparing performance metrics across all six events. Table~\ref{tab:post_eval} presents the performance of the N-BEATS model in predicting the onset of the equilibrium line, with most deviations remaining below one hour. E3 exhibits a slightly larger deviation of -1.17 hours, while a significant discrepancy of -4.765 hours is observed in E6. Because our system continuously generates real-time forecasts, we track hourly predictions until equilibrium is reached. Table~\ref{tab:post_eval_six} shows how forecast accuracy improves as the actual equilibrium time approaches.

\section{Conclusion}

In this work, we explored the application of state-of-the-art temporal learning models to forecast the start of equilibrium in tritium source activity for the KATRIN experiment. Accurate prediction of equilibrium start times is beneficial for optimizing measurement campaigns and improving operational efficiency.

We evaluated several models, including LSTM, N-BEATS, TFT, NHITS, DLinear, NLinear, TSMixer, and Chronos, on a datasets comprising 33 instability events. The models were assessed using metrics such as MAE, sMAPE, and DTW, along with additional criteria like execution time, prediction reliability, and moving time-step performance. Our analysis revealed that while simpler models (DLinear and NLinear) and LSTM failed to capture the declining trends, more advanced models like N-BEATS, TFT, NHITS, and TSMixer demonstrated strong predictive capabilities. Among these, N-BEATS emerged as the most suitable model due to its superior performance in curve structure capture (low DTW), repeatability (low standard deviation), and reliability in moving time-step predictions.

We also tested the hypothesis that smoothing the training datasets might enhance model performance. Contrary to our expectation, training on noisy datasets outperformed smoothed datasets, highlighting the importance of preserving inherent variability in the data for this use case.

Since other models also performed well, we could consider integrating several models into the monitoring system as part of an ensemble approach. In some scenarios, different models might offer better predictions, and by comparing the outputs of multiple models, we can increase confidence in predictions when they align closely. This ensemble strategy would enable more robust and reliable forecasting, particularly in cases where model performance may vary due to the characteristics of the dataset.

To enable real-world deployment, we integrated the N-BEATS model into the BORA framework, a lightweight and flexible data monitoring tool at KATRIN. This deployment allows the prediction framework to operate seamlessly in a real-time environment with minimal computational resources, leveraging pre-trained models to provide scientists with actionable insights during measurement campaigns.

This work demonstrates how advanced temporal learning models can address challenges in forecasting equilibrium conditions in large-scale experiments. Beyond the immediate application at KATRIN, the methods and approaches developed here have broader applicability in domains requiring time-series forecasting, particularly for scenarios involving noisy and irregular datasets. By automating the detection and prediction of critical events, this work sets the stage for more efficient experimental designs and data-driven decision-making processes.

\section*{Acknowledgment}

We gratefully acknowledge the support of the KATRIN experiment. We also appreciate the valuable contributions of Svenja Heyns and Rudolf Sack in reviewing the data analysis and manuscript. We thank Marco R\"ollig, Beate Bornschein, and Magnus Schl\"osser for approving the release of the BIXS count rate datasets used in this study. This work was supported by Helmholtz AI computing resources (HAICORE) through the Helmholtz Association’s Initiative and Networking Fund.

\bibliographystyle{IEEEtran}
{\footnotesize

}

\end{document}